\documentclass[aps,prl,twocolumn,showpacs,superscriptaddress]{revtex4}
\usepackage{graphicx}
\begin{document}

\title{Orbital Symmetry and Electron Correlation in Na$_{x}$CoO$_2$}
\author{W. B. Wu} \affiliation{National Synchrotron Radiation Research Center,
Hsinchu 30077, Taiwan}\affiliation{Department of Electrophysics,
National Chiao-Tung University, Hsinchu 300, Taiwan}
\author{D. J. Huang}
\altaffiliation[Corresponding author:] {\emph{
djhuang@nsrrc.org.tw}} \affiliation{National Synchrotron Radiation
Research Center, Hsinchu 30077, Taiwan} \affiliation{Department of
Electrophysics, National Chiao-Tung University, Hsinchu 300,
Taiwan}
\author{J. Okamoto}
\affiliation{National Synchrotron Radiation Research Center,
Hsinchu 30077, Taiwan}

\author{A. Tanaka}
\affiliation{Department of Quantum Matters, ADSM, Hiroshima
University, Higashi-Hiroshima 739-8530, Japan}

\author{H.-J. Lin}
\affiliation{National Synchrotron Radiation Research Center,
Hsinchu 30077, Taiwan}

\author{F. C. Chou}
\affiliation{Center for Materials Science and Engineering,
Massachusetts Institute of Technology, Cambridge, MA 02139,
U.S.A.}
\author{A. Fujimori}
\affiliation{Department of Complexity Science and Engineering,
University of Tokyo, Chiba 277-8561, Japan}
\author{C. T. Chen}
\affiliation{National Synchrotron Radiation Research Center,
Hsinchu 30077, Taiwan}

\date{\today }
\begin{abstract}
Measurements of polarization-dependent soft x-ray absorption
reveal that the electronic states determining the low-energy
excitations of Na$_{x}$CoO$_2$ have predominantly $a_{1g}$
symmetry with significant O $2p$ character. A large transfer of
spectral weight observed in O $1s$ x-ray absorption provides
spectral evidence for strong electron correlations in the layered
cobaltates. Comparing Co $2p$ x-ray absorption with calculations
based on a cluster model, we conclude that Na$_{x}$CoO$_2$
exhibits a charge-transfer electronic character rather than a
Mott-Hubbard character.
\end{abstract}

\maketitle Sodium cobalt oxides (Na$_{x}$CoO$_2$) have attracted
renewed interest because of their exceptionally large
thermoelectric power \cite{Terasaki97} and the recent discovery of
superconductivity in their hydrated counterparts \cite{Tekada03}.
Despite intensive experimental
\cite{Tekada03,Terasaki97,Ray99,Wang03,Motohashi03,Chainani03,Foo04,
Hassan04,Yang04} and theoretical
\cite{Singh00,Singh03,Q_Wang04,Marianetti04,Baskaran03,Koshibae03,Tanaka03,Zou03,Kumar03,Yanase04}
work, there remain many unresolved issues concerning the
electronic structure of Na$_{x}$CoO$_2$.

\begin{figure}[tbp]
\includegraphics[width=8cm]{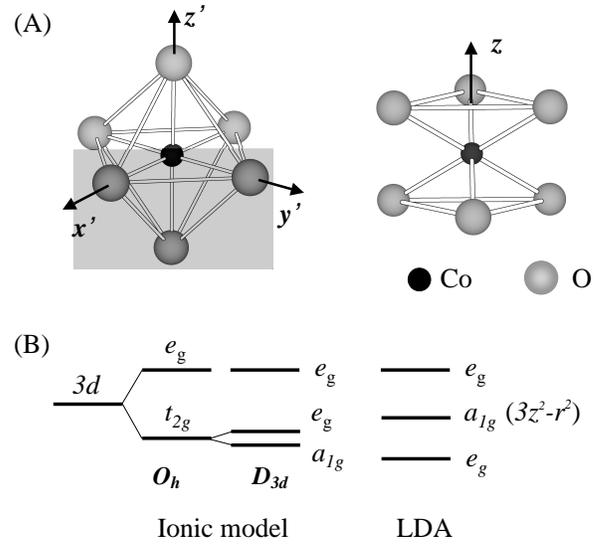}
\caption{\label{Fig_distortion} (A) Illustration of the trigonal
distortion of a CoO$_6$ octahedron. Left panel: undistorted
CoO$_6$ octahedron with cubic ($O_h$) symmetry. Right panel:
compressed CoO$_6$ octahedron with $D_{3d}$ symmetry. The
distorted CoO$_6$ is rotated such that the three-fold rotation
axis is along the $c$-axis. (B) Crystal-field splitting of Co $3d$
states in distorted CoO$_6$ according to an ionic model and
relative energy positions of $3d$ bands obtained from LDA
calculations.}
\end{figure}

An important issue is the orbital character of the valence
electrons responsible for low-energy excitations. The lattice of
Na$_{x}$CoO$_2$ exhibits a trigonal distortion; its CoO$_2$ layer
consists of a triangular net of distorted edge-sharing oxygen
octahedra with Co at the center, leading to a splitting of
$t_{2g}$ states into $e_{g}^{\pi}$ and $a_{1g}$ states as shown in
Fig. \ref{Fig_distortion}. The $e_{g}^{\pi}$ states spread over
the $ab$ plane, whereas the $a_{1g}$ state extends to the
$c$-axis; in a coordination system in which the $z$-axis is along
the $c$-axis, the $a_{1g}$ state is $d_{3z^{2}-r^{2}}$ and the two
$e_g^{\pi}$ states are $\frac{1}{\sqrt{3}}(d_{yz}+\sqrt{2}d_{xy})$
and $\frac{1}{\sqrt{3}}(d_{zx}-\sqrt{2}d_{x^{2}-y^{2}})$.
Band-structure calculations in the local-density approximation
(LDA) show that the $a_{1g}$ state at the $\Gamma$ point has a
one-particle energy of 1.6~eV higher than that of the
$e_{g}^{\pi}$ states \cite{Singh00}. Thereby, the $e_{g}^{\pi}$
states are almost filled, while the $a_{1g}$ is partially filed
and is the electronic state most relevant to low-energy
excitations. These calculations are however different from a
viewpoint based on a crystal-field approach. According to a
point-charge model, one would expect that the compressed trigonal
distortion stabilizes the $a_{1g}$ state with an energy $\sim$
0.025~eV lower than that of $e_{g}^{\pi}$ \cite{Koshibae03}, as
illustrated in Fig. \ref{Fig_distortion}(B).

To comprehend the effect of electron correlations and the trigonal
distortion on $t_{2g}$ states are imperative for an understanding
of the electronic structure of Na$_{x}$CoO$_2$. One requires
spectral evidence for electron correlations to justify a
microscopic model of correlated electrons to explain the
spectacular physical properties of Na$_{x}$CoO$_2$. Based on the
LDA results \cite{Singh00} that the $a_{1g}$ band is separated
from the $e_{g}^{\pi}$ bands, authors of several theoretical works
have proposed one-band models for the superconductivity, such as
the extended Hubbard model \cite{Tanaka03} and the $t-J$ model
based on the resonating valence bond (RVB) theory
\cite{Q_Wang04,Baskaran03,Kumar03}. In contrast, multi-band models
\cite{Koshibae03,Yanase04} were also proposed for the triangular
cobaltates. Thus evidence for electron correlations and the
validity of one-band models are fundamental questions for the
underlying physics of Na$_{x}$CoO$_2$.

Another primary issue is whether the electronic states of
Na$_{0.5}$CoO$_2$ responsible for low-energy excitations have O
$2p$ character. Early LDA calculations \cite{Singh00} indicate
that hybridization between Co $3d$ and O $2p$ in Na$_{0.5}$CoO$_2$
is weak. Analysis of core-level photoemission results suggests
that Na$_{x}$CoO$_2$ has a Mott-Hubbard-like rather than a
charge-transfer electronic structure \cite{Chainani03}. On the
other hand, recent LDA results corroborated with a Hubbard-like
model conclude that Na$_{1/3}$CoO$_2$ exhibits significant
hybridization between Co $3d$ and O $2p$ states
\cite{Marianetti04}.

In this Letter, we present measurements of soft x-ray absorption
spectra (XAS) on Na$_{x}$CoO$_2$ pertinent to its electronic
structure such as orbital character of the electronic states
determining the low-energy physics. In addition, we discuss the
spectral character of strongly correlated electrons in a
one-electron addition process. Comparison of O 1s XAS with various
doping shows that Co $3d$ electrons are strongly correlated. To
investigate the detailed electronic structure, we compared Co $2p$
XAS with results of calculations using a cluster model in the
configuration-interaction (CI) approach.

Single crystals of Na$_{0.75}$CoO$_2$ were grown by the traveling
solvent floating-zone method. Crystals with smaller Na
concentrations of x = 0.67 and 0.5 were prepared from
Na$_{0.75}$CoO$_2$ through subsequent electrochemical
de-intercalation procedures, as confirmed by Electron Microprobe
Analysis. Details of crystal growth, electrochemical
de-intercalation, and characterization of the resulting samples
are discussed elsewhere \cite{Chou04}.

We measured XAS on Na$_{x}$CoO$_2$ single crystals using the
Dragon beamline at the National Synchrotron Radiation Research
Center in Taiwan. The XAS were recorded through collecting the
sample drain current. Crystals were freshly cleaved in ultra-high
vacuum with a pressure lower than $5\times 10^{-10}$ torr at 80~K.
The incident angle was $60^{\circ}$ from the sample surface
normal; the photon energy resolutions were set at 0.12~eV and
0.25~eV for photon energies of 530~eV and 780~eV, respectively. In
polarization-dependent measurements, we rotated the sample about
the direction of incident photons to obtain XAS from which
experimental artifacts related to the difference in the optical
path and to the probing area have been eliminated \cite{Huang04}.
All measured XAS referred to the \textbf{E} vector parallel to the
$c$-axis are shown with a correction for the geometric effect,
$I_{\parallel}=\frac{4}{3}(I-\frac{1}{4}I_{\perp})$, in which
$I_{\perp}$ and $I$ are measured XAS intensities with
\textbf{E}~${\perp}~c$ and with \textbf{E} in the plane defined by
the $c$-axis and the direction of incident radiation,
respectively; $I_{\parallel}$ is the deduced XAS intensity for
\textbf{E}~${\parallel}~c$.

\begin{figure}[tbp]
\includegraphics[width=8cm]{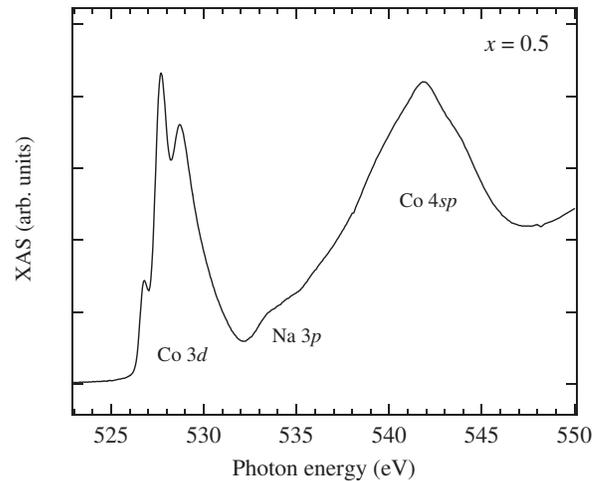}
\caption{\label{Fig_O1s_w}O $1s$ XAS spectrum of Na$_{0.5}$CoO$_2$
measured in the total electron yield mode. The unoccupied bands
with which O $2p$ states hybridize are denoted in the plot.}
\end{figure}

Figure \ref{Fig_O1s_w} presents the O $1s$ XAS of
Na$_{0.5}$CoO$_2$. O 1s XAS measures transitions from an O $1s$
core level to unoccupied O $2p$ states mixing with bands of
primary Co or Na character. One can interpret O $1s$ XAS as a
one-electron addition process, i.e., $d^{n}\rightarrow d^{n+1}$,
if the influence of the O $1s$ core hole is neglected
\cite{Groot89,Elp99}. The structure in our measured XAS near the
threshold arises from covalent mixing of Co $3d$ and O $2p$; the
broad feature about 540 eV corresponds to Co $4sp$ bands. The
features in the region about 535 eV are attributed to transitions
involving Na $3p$. We observed three pronounced O 1s XAS peaks in
the vicinity of the absorption threshold, implying strong
hybridization between O $2p$ and Co $3d$ and many O $2p$ holes
existing in Na$_{0.5}$CoO$_2$.

\begin{figure}[tbp]
\includegraphics[width=8cm]{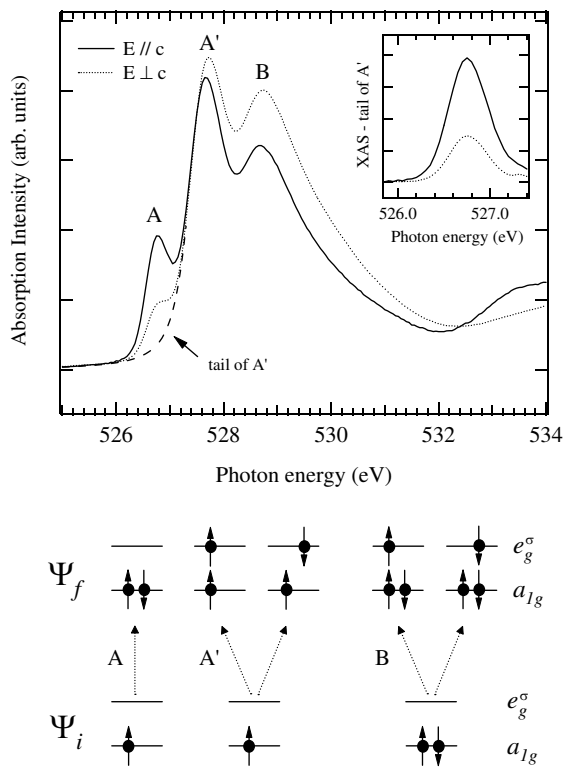}
\vspace{-2mm} \caption{\label{Fig_O1s} Upper panel:
Polarization-dependent O 1s XAS of Na$_{0.5}$CoO$_2$ with the
\textbf{E} vector of the light perpendicular (dotted line) and
parallel (solid line) to the $c$-axis. The inset shows the XAS of
peak A after removal of background resulting from the tail of peak
A$'$ (dashed line). Lower-panel: Energy diagrams illustrating
transitions from $\Psi_{i}$ in the low-spin state to $\Psi_{f}$
corresponding to the symmetries of peaks A, A$'$ and B labelled in
the upper panel. The fully occupied $e_{g}^{\pi}$ states are
omitted from the energy diagrams for clarity.} \vspace{-2mm}
\end{figure}

As for the symmetry of the pre-peaks corresponding to Co $3d$
bands, we resorted to measurements of polarization-dependent O
$1s$ XAS. Figure \ref{Fig_O1s} depicts O $1s$ XAS of
Na$_{0.5}$CoO$_2$ with the \textbf{E} vector of photons
perpendicular ($I_{\perp}$) and parallel ($I_{\parallel}$) to the
crystal $c$-axis. The O $1s$ XAS shows that a peak at 526.8 eV
(labelled as A) has a strong $z$ component, in contrast to the
in-plane orbital $d_{x^{2}-y^{2}}$ in the cuprates \cite{Chen92}.
The ratio $I_{\perp}/I_{\parallel}$ for peak A is 0.37~$\pm$~0.05,
as shown in the inset of Fig. 3. The in-plane components of two
other peaks at 526.8 eV and 527.6 eV (labelled as A$'$ and B,
respectively) are slightly larger than their corresponding $z$
components.

The relative intensities of peaks A, A$'$ and B in the O $1s$ XAS
depend on hybridization between Co $3d$ and O $2p$. Qualitatively,
the hybridization results from the inter-atomic matrix element
$V_{pd}$ between Co $3d$ and O $2p$, which can be expressed in
terms of the Slater-Koster transfer integrals $pd\sigma$ and
$pd\pi$ \cite{Slater54}; the ratio $I_{\perp}/I_{\parallel}$ of O
$1s$ XAS is proportional to the ratio of the averaged $V_{pd}^2$
with O 2$p$ orbitals perpendicular and parallel to the $c$-axis.
For an undistorted lattice, $I_{\perp}/I_{\parallel}$ of O $1s$
XAS with a final state of $a_{1g}$ symmetry is 0.25, while that of
$e_{g}$ symmetry is 1.0, if one uses an empirical relation
$pd\sigma=-(4/\sqrt{3})pd\pi$. $I_{\perp}/I_{\parallel}$ depends
also on the distortion and the band effect. If the compressed
trigonal distortion of a Na$_{x}$CoO$_2$ lattice is taken into
account, O 1$s$ XAS with final states of $a_{1g}$ symmetry has a
large out-of-plane polarization, whereas that with $e_g$ symmetry
has an in-plane polarization. Thus, our measurements that peak A
and A$'$ have opposite polarizations show that peak A results
predominantly from adding an electron to a state of $a_{1g}$
symmetry, whereas peaks A$'$ and B correspond to adding electrons
to states of $e_{g}$ symmetry, as illustrated in the lower panel
of Fig. \ref{Fig_O1s}. In other words, the symmetries of the
transitions associated with peaks A, A$'$ and B correspond to
$(a_{1g})^{1}{\rightarrow}(a_{1g})^{2}$,
$(a_{1g})^{1}{\rightarrow}(a_{1g})^{1}(e_{g}^{\sigma})^{1}$, and
$(a_{1g})^{2}{\rightarrow}(a_{1g})^{2}(e_{g}^{\sigma})^{1}$,
respectively. Note that $(e_{g}^{\pi})^4$ is omitted from above
expressions of one-electron addition for clarity. These
observations reveal that the electronic states determining the
low-energy excitations of Na$_{x}$CoO$_2$ have predominantly
$a_{1g}$ symmetry, demonstrating that one-band models for the
superconductivity of hydrated Na$_{x}$CoO$_2$ is a reasonable
approach. Moreover, our observation of the $a_{1g}$ symmetry of
states crossing the Fermi level suggests a significant
hybridization between O $2p$ and Co $3d$, because hopping of
$a_{1g}$ states within the CoO$_2$ layer would be difficult
without O $2p$ mixing. Hence, one expects that the ground-state
configurations for Co$^{4+}$ and Co$^{3+}$ in Na$_{x}$CoO$_2$ have
significant weights of $d^{6}\underline{L}$ and
$d^{7}\underline{L}$, respectively, in which $\underline{L}$
denotes an oxygen $2p$ hole.
\begin{figure}[tbp]
\includegraphics[width=8cm]{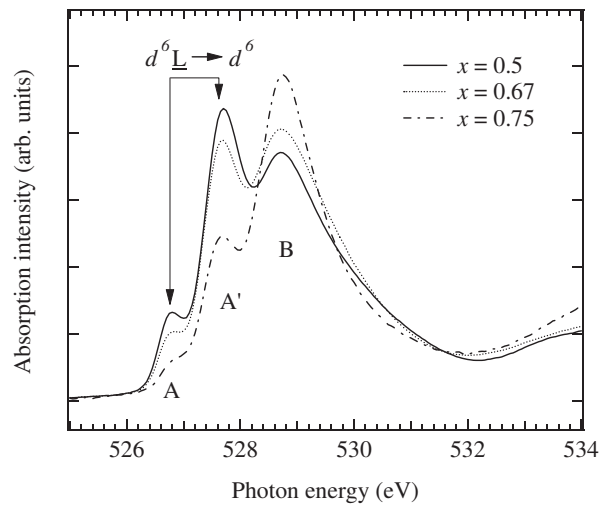}
\caption{\label{Fig_O1s_doping} Doping-dependent isotropic O $1s$
XAS of Na$_{x}$CoO$_2$, i.e., $(I_{\parallel}+I_{\perp})/2$. The
dominant transitions $d^{6}\underline{L}$~$\rightarrow$~$d^6$
leading to the peaks A and A$'$ are indicated in the figure.}
\end{figure}

To seek spectral evidence for electron correlations of $3d$ bands,
we plot doping-dependent isotropic O $1s$ XAS of Na$_{x}$CoO$_2$
in Fig. \ref{Fig_O1s_doping}. The spectra are normalized to have
the same intensity at 600 eV at which O $1s$ XAS has no
doping-dependent structure. Doping of Na donates electrons to the
hypothetical mother compound CoO$_2$ and changes a fraction $x$ of
Co$^{4+}$ to Co$^{3+}$. Remarkably, we found that, as the doping
$x$ increases, the intensities of peaks A and A$'$ decrease, but
peak B increases in intensity. Thus change in the spectral weight
of O $1s$ XAS indicates that peaks A and A$'$ and peak B arise
from transitions of adding an electron to the unoccupied states on
the Co$^{4+}$ and Co$^{3+}$ sites, respectively. Such a spectral
weight transfer is in contrast to the prediction of a one-electron
theory in which a rigid-band shift is expected and the spectral
weight of $3d$ bands is influenced only by the position of the
Fermi level. The spectral-weight transfer of the one-electron
addition observed in Na$_{x}$CoO$_2$ is a general feature of
strongly correlated systems \cite{Eske91}, as in electron-energy
loss experiments \cite{Romberg90} and O $1s$ XAS \cite{Chen91}
study of La$_{2-x}$Sr$_{x}$CuO$_4$. Such behavior has been found
in Li-doped NiO as well \cite{Kuiper89}. Both
La$_{2-x}$Sr$_{x}$CuO$_4$ and Li$_{x}$Ni$_{1-x}$O are
charge-transfer systems; thus Na$_{x}$CoO$_2$ is expected to have
a charge-transfer electronic structure.

\begin{figure}[tbp]
\includegraphics[width=8cm]{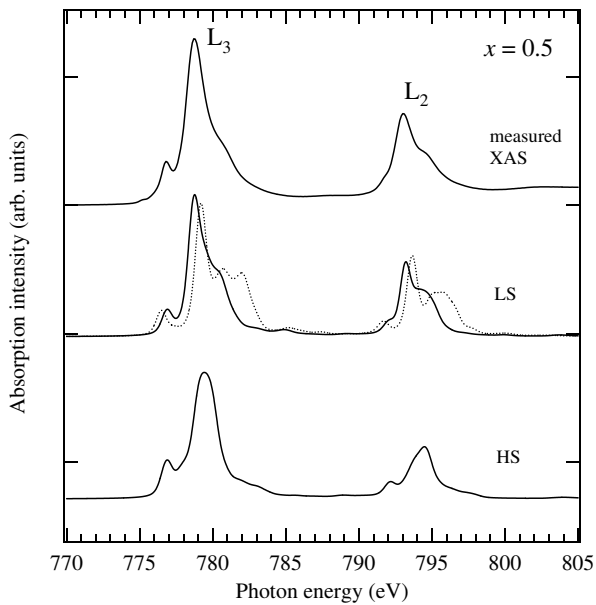}
\vspace{-3mm} \caption{\label{Fig_Co2p} Measured and calculated
isotropic Co L-edge XAS of Na$_{0.5}$CoO$_2$. Theoretical Co
L-edge XAS spectra for LS and HS Co $3d$ states were broadened
with a Gaussian full width of 0.25 eV at half maximum (FWHM) and
with a Lorentian FWHM of 0.2 eV. The dotted line is a simulated
XAS with the parameters reported in Ref. \cite{Chainani03}.}
\end{figure}

We measured also Co L$_{2,3}$-edge XAS of Na$_{x}$CoO$_2$ to
further study its detailed electronic structure. Based on a
cluster model in the CI approach \cite{Chen04}, we simulated XAS
spectra with a superposition of calculated XAS for Co$^{4+}$ and
Co$^{3+}$ with weights of $1-x : x$. Details of the calculations
will be presented elsewhere. To summarize, the shoulder peak on
the low-energy side of the L$_3$ edge results from the $a_{1g}$
orbital character in the ground state. Comparing the Co L-edge XAS
of $x$~=0.5, 0.67, and 0.75 with CI calculations using a series of
parameters, we found that the calculated XAS for Co ions in a
low-spin (LS) state resembles the measured XAS satisfactorily, but
the calculated XAS for high-spin (HS) ions is inconsistent with
the measurement, as demonstrated in Fig. \ref{Fig_Co2p} for
$x=0.5$ \cite{CI}. The calculations indicate that Na$_{x}$CoO$_2$
has a charge-transfer energy smaller than the on-site Coulomb
energy ($U_{dd}$=4.5 eV). In particular, because of the high
valency, Co$^{4+}$ ions have a negative charge-transfer energy
(${\Delta}\sim -1$~eV) \cite{Delta}, in contrast to the conclusion
from the analysis of core-level photoemission data
\cite{Chainani03}. Calculations with the parameters $U_{dd}$=5.5,
$\Delta$=3.1, $10Dq=2.5$ for Co$^{3+}$, and $10Dq=4.0$ for
Co$^{4+}$, in units of eV, that Chainani \emph{et al.}
\cite{Chainani03} concluded, give rise to a Co $2p$ XAS (dotted
line in Fig. \ref{Fig_Co2p}) inconsistent with our measurements.
Thus Na$_{x}$CoO$_2$ exhibits a charge-transfer electronic
character rather than a Mott-Hubbard character; the
$d^{6}\underline{L}$ configuration dominates the ground state of
Co$^{4+}$ in Na$_{x}$CoO$_2$ \cite{CI1}, like Co$^{4+}$ in
SrCoO$_3$ \cite{Potze95} and La$_{1-x}$Sr$_{x}$CoO$_3$
\cite{Saitoh97}. These results suggest that peaks A and A$'$ of
Fig. \ref{Fig_O1s_doping} are derived from transitions of
$d^{6}\underline{L}$~${\rightarrow}$~$d^6$ and that the empty
$a_{1g}$ band to which the doped electrons go has predominantly O
$2p$ character.

In conclusion, measurements of doping-dependent O $1s$ XAS provide
a spectral fingerprint for strong correlations of $3d$ electrons
in doped 2D triangular cobalt oxides. Our results reveal the
charge-transfer electronic character of Na$_{x}$CoO$_2$; the
doping of Na strongly affects the O $2p$ hole density. The
electronic states responsible for the low-energy excitations of
Na$_{x}$CoO$_2$ have predominantly $a_{1g}$ symmetry with
significant O $2p$ character.

\begin{acknowledgements}
We thank F. C. Zhang, C. Y. Mou, G. Y. Guo, and L. H. Tjeng for
valuable discussions. This work was supported in part by the
National Science Council of Taiwan and by the MRSEC Program of the
National Science Foundation of U.S. under award number
DMR-02-13282.

\end{acknowledgements}


\begin{thebibliography}{99}
\bibitem{CI1} The ground-state configurations for Co$^{4+}$
are $d^5$=15.9\%, $d^{6}\underline{L}$=48.1\%,
$d^{7}\underline{L}^2$=31.1\%, and $d^{8}\underline{L}^3$=4.9\%;
those for Co$^{3+}$ are $d^6$=43.2\%, $d^{7}\underline{L}$=46.9\%,
and $d^{8}\underline{L}^2$=9.9\%.
\end{thebibliography}
\end{document}